\documentclass[a4paper,11pt]{article}
\usepackage{jheppub}
\usepackage[T1]{fontenc}
\usepackage{float}
\usepackage{graphicx}
\usepackage{graphics}
\usepackage{dcolumn}   
\usepackage{color}
\usepackage[english]{babel} 
\usepackage{psfrag}
\usepackage{tabularx}
\usepackage{bbm,bm}
\usepackage{appendix}
\numberwithin{equation}{section}
\usepackage{rotating}
\usepackage{epsfig}
\def\thesection{\arabic{section}}
\def\thesubsection{\arabic{section}.\arabic{subsection}}
\def\thesubsubsection{\arabic{section}.\arabic{subsection}.\arabic{subsubsection}}
\usepackage{hyperref}
\allowdisplaybreaks

\usepackage{physics}

\begin{document}
%


\title{Pulsating string solution and stability in two parameter $\chi$-deformed background}
\author{Rashmi R Nayak$^1$,}
\author{Nibedita Padhi$^2$,}
\author{Manoranjan Samal$^3$}
\affiliation{$^1$Centre for Ocean, River, Atmosphere and Land Sciences, Indian Institute of Technology Kharagpur, Kharagpur 721 302, India}
\affiliation{$^2$Department of Physics,Indian Institute of Technology Kharagpur, Kharagpur 721 302, India}
\affiliation{$^3$Rayagada Autonomous College, Rayagada, 765001, India}
\emailAdd{rashmi@coral.iitkgp.ac.in} \emailAdd{nibedita.phy@iitkgp.ac.in}
\emailAdd{manoranjan.phys@gmail.com}
 
\abstract{In this paper, we investigate pulsating string solutions within the context of a two-parameter $\chi$-deformed $\mathcal{R} \times S^2$ background. We derive the energy and  oscillation number relation for the deformed $\mathcal{R} \times S^2$ in a short string limit. Additionally, we derive the first-order perturbation equation for a pulsating string in this specific background. As a special case of the two-parameter deformed background, we examine the stability of the pulsating string solution in  one-parameter deformed $\mathcal{R}  \times S^2$ and $AdS_3$ background by setting one of the deformation parameters to zero.}

\maketitle

\section{Introduction}
A remarkable example of AdS/CFT correspondence lies in the complex connection between the spectrum of close superstring in $AdS_5 \times S^5$ space time and gauge invariant operators in 4d $\mathcal{N}=4$ Supersymmetric Yang-Mills theory \cite{Maldacena:1997re}. A precise match between the string states and the dual operators of both the sides is hard to achieve. However, it has been shown that in the plane wave limit it is possible to identify the energy of the string states with the anomalous dimension of the gauge invariant operators considering large R-charge \cite{Berenstein:2002jq, Santambrogio:2002sb}. There have been a lot of studies in various settings over the years using a variety of classical strings as probes, such as folded strings \cite{Gubser:2002tv}, giant magnons \cite{Hofman:2006xt}, spiky strings \cite{Kruczenski:2004wg} along with their detailed dual gauge theories. Pulsating strings are distinct from all other string solutions due to their time-dependent nature and better stability. In Ref. \cite{Minahan:2002rc} the dimension of the dual operator has been predicted by computing the energy of circular string pulsating in $AdS_5 \times S^5$ with semiclassical limit. Similar analysis has also led to extensive work on pulsating strings in different backgrounds, where the energy of the string states can be expressed as complicated functions of string tension and various charges \cite{Khan:2003sm, Dimov:2004xi, Smedback:2004udl, Kruczenski:2004cn, Chen:2008qq, Dimov:2009rd, Bobev:2004id, Arnaudov:2010by, Park:2005kt, Pradhan:2013sja, Beccaria:2010zn, Giardino:2011jy, Banerjee:2014bca, Panigrahi:2014sia, Rotating2, Barik1, Banerjee:2014rza, HN3, Banerjee:2016xbb,Chakraborty:2022eeq}. Recently, we have studied a pulsating string in  NS5 brane and I-brane background \cite{Biswas:2023uuq}, by expressing its energy in terms of the adiabatic invariant $\mathcal{N}$. In this article, we will consider a multiwrapped pulsating string moving in a two-parameter deformed $\mathcal{R} \times S^2$ background. \medskip

Understanding the stability properties of such string solutions on curved space backgrounds is also a topic of great interest. One effective approach to comprehend the stability properties is by analysing the worldsheet fluctuations of string in curved backgrounds. The authors of Ref. \cite{Larsen:1993iva} have developed a covariant approach to study the dynamics of the small disturbances. The procedure involves performing a perturbative series expansion around the precisely known solution for a generic string in the specified curved spacetime. Following the covariant approach, it has been investigated the emergence of string instabilities in D-dimensional black hole spacetimes and de-Sitter spacetime. 
The perturbation of spiky strings in flat and AdS space has been studied in  \cite{Bhattacharya:2016ixc, Bhattacharya:2018unr, Bhattacharya:2021xfc}, a circular string in a power law expanding universe was shown in \cite{Larsen:1994jt}, the planetoid strings in Ellis geometry and in (2+1) BTZ black hole is discussed in \cite{Kar:1997zi} and a pulsating string in $AdS_3 \times S^3$ was studied in \cite{Barik:2017opb}. 
Following this method numerous studies have been conducted in various gravitational and cosmological backgrounds, one can check \cite{Garriga:1991ts, Guven:1993ew, Guven:1993ex, Larsen:1994ah, Larsen:1993mx, Khan:2005fc, Larsen:2003ma, Larsen:1996yb, Mahapatra:1996ap, Larsen:1995af,  Li:2018jxy, Norma:2016soj, Hioki:2009hp, Brandenberger:2007by}. The covariant formalism of perturbation has also been used to understand the stability properties of moving branes.  A higher order perturbation equation for branes has been formulated in \cite{Capovilla:2021xfy} and first order perturbation equation for a macroscopic string and relativistic membranes moving in an arbitrary curved space-time was derived following the covariant formalism in \cite{Larsen:2000sg, deVega:1988ch, Kiosses:2014tua}. In addition to the above-mentioned works, semiclassical quantization makes use of the study of tiny perturbations surrounding classical string solutions. Numerous studies have also been made in this direction. Originally Vega and Sanchez developed the method to study string solutions in the presence of a strong curvature regime \cite{deVega:1987veo}. This method computes first and second order quantum fluctuations, which result in the determination of physical quantities such as scattering amplitudes and mass spectrum. For additional details on this one can check  \cite{deVega:1987veo, deVega:1988jh}. The purpose of the present paper is to study the stability of small perturbations on pulsating strings moving in a single and two parameter  $\chi$ deformed $\mathcal{R} \times S^2$ background.\medskip

Regarding the deformed backgrounds, the construction of one parameter deformation of the Matsaev- Tseytlin action in the context 
of supercoset sigma models describe dynamics
of a system in terms of fields, leading to classically integrable model \cite{Delduc:2013qra}. This construction
involves q-deformed symmetries, spacetime metric deformations and the development of
a R-matrix with associate S-matrix and dispersion relation for certain string theories in specific spacetime backgrounds \cite{Beisert:2008tw, Hoare:2023zti, Hoare:2022asa}.
On the other hand, a two parameter deformed sigma model on the world sheet of superstrings on $AdS_3\times S^3 \times M^4$ was first constructed by Hoare in \cite{Hoare:2014oua}. The deformed metric involves two parameters $\chi_{+}$ and $\chi_{-}$ and due to the presence of a $\textbf{Z}_2$ symmetry of the geometry $\chi_{+}$ $\leftrightarrow$ $\chi_{-}$, we can take  $\chi_{+}$ ~$\geq $ ~$\chi_{-}$  without any loss of generality. Despite the deformation, the classical integrability of the original model is preserved. The metric  reduces to one-parameter deformed case  \cite{Hoare:2014pna} taking $\chi_{+} \rightarrow 0$  and to a known squashed $S^3$  case \cite{Cherednik:1981df} for  $\chi_{+}$ ~$=$ ~$\chi_{-}$. The formula for Ramond-Ramond fluxes in the two parameter deformed background has been formulated in \cite{Seibold:2019dvf} and the authors of \cite{Seibold:2021lju} has computed the massive tree level scattering matrices for the two parameter deformed background using perturbation theory. A quantum spin chain model on this background has been studied in \cite{Wen:2019mgv}. The authors have also found the complicated dispersion relation for giant magnon solutions and various hanging string solutions treating the deformation parameters perturbatively.\medskip

In order to understand the interesting aspect of $\chi$-deformed sigma models we have considered the pulsating string solutions with two parameter deformation in the background of $\mathcal{R} \times S^2$ in section 2. In the first case, we study the string motion which seems to be equivalent to the motion of a particle under the influence of a potential $V(\theta)$ for different values of $\chi_{+}$ and $\chi_{-}$. Then we have used the Bohr-Sommerfeld like quantisation to find the oscillation number $\mathcal{N}$ and its relation to the string energy, whereas the one parameter case has been studied in \cite{Banerjee:2016xbb}. In section 3 we have studied the propagation of a perturbation along the pulsating string and derived the equation governing perturbation. We have also checked the stability of such string solutions in the deformed background where one of the deformation parameter is set to zero. Finally, in section 4 we present our conclusion.


\section{Pulsating string in two parameter deformed $\mathcal{R} \times S^2$}
In this section, we will delve into the semiclassical quantisation of a closed pulsating string in two parameter $\chi$ deformed $\mathcal{R} \times S^2$. The generic metric for a two parameter deformed $AdS_3\times S^3$ background is given by \cite{Hoare:2014oua},
\begin{equation}
    \begin{split}
        ds^2=&\frac{1}{F(\rho)}\left[-(1+\rho^2)dt^2+\frac{d\rho^2}{1+\rho^2}+\rho^2d\psi^2-\left(\chi_-(1+\rho^2)dt-\chi_+\rho^2d\psi\right)^2\right]\\&+\frac{1}{\Tilde{F}(r)}\left[(1-r^2)d\zeta^2+\frac{dr^2}{1-r^2}+r^2d\phi^2-\left(\chi_-(1-r^2)d\zeta+\chi_+r^2d\phi\right)^2\right],
        \label{metric1}
           \end{split}
\end{equation}
with $$F(\rho)=1+\chi_-^2(1+\rho^2)-\chi_+^2\rho^2~,~~\Tilde{F}(r)=1+\chi_-^2(1-r^2)+\chi_+^2r^2.$$
The metric is not supported by any B- field as it is a total derivative. We now consider $\rho=0$ and $r=\sin{\theta}$ in equation (\ref{metric1}) which reduces it to the deformed $\mathcal{R} \times S^2$  background
\begin{equation}
\begin{split}
    ds^2=\frac{1}{\Tilde{F}(\theta)}&\bigg[-\Tilde{F}(\theta)dt^2+\cos^2{\theta}(1+{\chi_-}^2\cos^2{\theta})d\zeta^2+d\theta^2+\sin^2{\theta}(1+\chi_+^2\sin^2{\theta})d\phi^2\\&~~~~~+2\chi_-\chi_+\cos^2{\theta} \sin^2{\theta}d\phi d\zeta\bigg], \\& \text{where}~~~~\Tilde{F}(\theta)=1+\chi_-^2cos^2{\theta}+\chi_+^2sin^2{\theta}.
    \label{2.2}
\end{split}
\end{equation}
As our interest lies in a fundamental string motion, we consider the Polyakov action  
\begin{eqnarray}
    S_{P}=-\frac{\sqrt{\hat{\lambda}}}{4\pi}\int d\tau d\sigma \sqrt{-h}h^{\alpha \beta}G_{\alpha\beta}.
    \label{polyakov}
\end{eqnarray}
where $\hat{\lambda}$ is the modified 't Hooft coupling for the deformed model, $h^{\alpha \beta}$ is the internal metric and  $\gamma^{\alpha \beta}$ is the worldsheet metric. The induced metric $G_{\alpha \beta}$ on the string worldsheet  is  defined by
\begin{equation}
    G_{\alpha \beta}=g_{\mu\nu}\frac{\partial X^{\mu}}{\partial \sigma^{\alpha}}\frac{\partial X^{\nu}}{\partial \sigma^{\beta}},
    \label{induced}
\end{equation}
where $\sigma^{\beta}$ are worldsheet coordinates and $X^{\mu}$ are space time coordinates.
The equations of motion derived from the action (\ref{polyakov}) along with the conformal gauge constraint are given by
\begin{equation}
    \Ddot{X^{\mu}}-X''^{\mu}+\Gamma^{\mu}_{\rho\sigma}\Big(\Dot{X^{\rho}}\Dot{X^{\sigma}}-X'{\rho}X'{\sigma}  \Big)=0\label{eom},
\end{equation}
and,
\begin{equation}
    G_{\alpha \beta}-\frac{1}{2}h_{\alpha \beta}G^{\gamma}_{~\gamma}=0.
    \label{gauge}
\end{equation}
We now  take the ansatz for a pulsating string in the deformed $\mathcal{R} \times S^2$ background (\ref{metric1}) as
\begin{equation} t=t(\tau),~~~\theta=\theta(\tau),~~\zeta=0,~~~\phi=m\sigma.  \label{ansatz}
\end{equation}
With the help of the ansatz the resulting action is given by
\begin{equation}
S=-\frac{\sqrt{\hat{\lambda}}}{4\pi}\bigg[\Dot{t}^2+\frac{1}{\Tilde{F}(\theta)}\left(m^2sin^2{\theta}(1+{\chi_+}^2sin^2{\theta})-\Dot{\theta}^2\right)\bigg].
\label{action}
\end{equation}
 The equations of motion for coordinate $t$  and $\theta $ give 
\begin{align}
\Dot{t}=\mathcal{E}, \label{teq}\\
2\Tilde{F}^2(\theta)\Ddot{\theta}+\dot{\theta}^2+m^2\sin{2\theta}(1+2\chi_{+}^2\sin^2{\theta})\Tilde{F}(\theta)-m^2\sin^2{\theta}(1+\chi_{+}^2\sin^2{\theta})=0,
\label{thetaeq}
 \end{align}
and the conformal gauge constraint equation (\ref{gauge}) gives
\begin{equation}
    \Dot{\theta}^2=(1+\chi_-^2\cos^2{\theta}+\chi_+^2\sin^2{\theta})\mathcal{E}^2-m^2\sin^2{\theta}(1+\chi_+^2\sin^2{\theta}).
    \label{virasoro}
\end{equation}
Equation (\ref{virasoro}) can also be written in the following form,  
\begin{equation}
    \Dot{z}^2=m^2\chi_+^2 (1-z^2)(z^2-Z_1)(Z_2-z^2),
    \label{eq}
\end{equation}
where $z=\sin{\theta}$ and  $Z_1$ and $Z_2$  are given by
\begin{align*}
       &Z_1=\frac{1}{2} \left(-\frac{1}{{\chi_+}^2}+\frac{\mathcal{E}^2 \left({\chi_+}^2-{\chi_-}^2\right)}{m^2
  {\chi_+}^2}-\sqrt{\left(\frac{1}{{\chi_+}^2}-\frac{\mathcal{E}^2 \left({\chi_+}^2-{\chi_-}^2\right)}{m^2{\chi_+}^2}\right)^2+\frac{4
   \mathcal{E}^2 \left({\chi_-}^2+1\right)}{m^2 {\chi_+}^2}}\right),\\
   & Z_2= \frac{1}{2} \left(-\frac{1}{{\chi_+}^2}+\frac{\mathcal{E}^2 \left({\chi_+}^2-{\chi_-}^2\right)}{m^2
  {\chi_+}^2}+\sqrt{\left(\frac{1}{{\chi_+}^2}-\frac{\mathcal{E}^2 \left({\chi_+}^2-{\chi_-}^2\right)}{m^2{\chi_+}^2}\right)^2+\frac{4
   \mathcal{E}^2 \left({\chi_-}^2+1\right)}{m^2 {\chi_+}^2}}\right).
 \end{align*}
Now, solving the differential equation (\ref{eq}) for $\dot{z}$ yields the solution of string motion in terms of Jacobi functions\footnote{$cn(x)$ is the solution to  differential equations~$$ y'(x)^2-(1-y(x)^2)(1-m+my(x)^2)=0$$, and $sn(x)$ is solution to the differential equations ~$$ y'(x)^2-(1-y(x)^2)(1-my(x)^2)=0$$} as
\begin{equation}
 \sin{\theta}=z(\tau)=\cfrac{\sqrt{Z_1 Z_2} ~sn \left(m \chi _{+}  \sqrt{Z_2-Z_1}\tau ~|~\frac{\left(Z_1-1\right)
   Z_2}{Z_1-Z_2}\right)}{\sqrt{Z_1-Z_2 ~ cn\left(m\chi _{+} \sqrt{Z_2-Z_1} \tau ~|~\frac{\left(Z_1-1\right) Z_2}{Z_1-Z_2}\right) ^2}}.
   \label{sol}
\end{equation}

\begin{figure}[ht!]
    \centering
\includegraphics[width=0.45\textwidth,height=4.5cm]{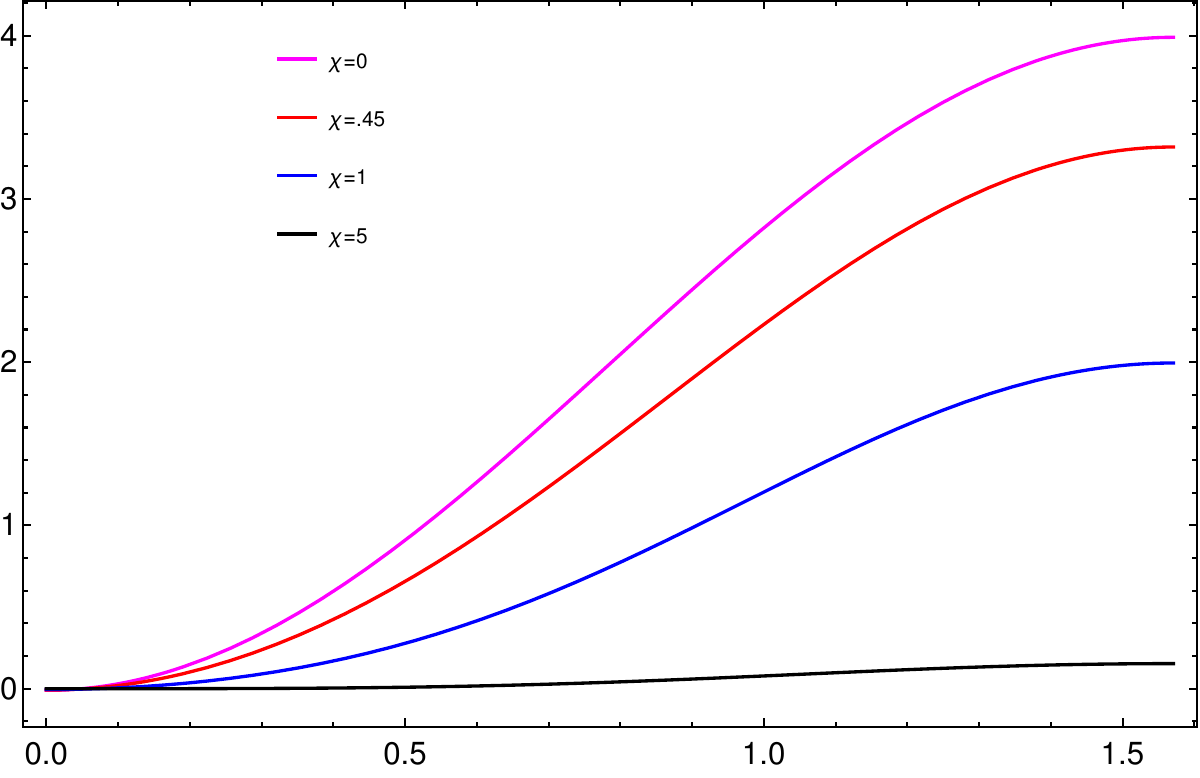}\hfill
\includegraphics[width=0.45\textwidth,height=4.5cm]{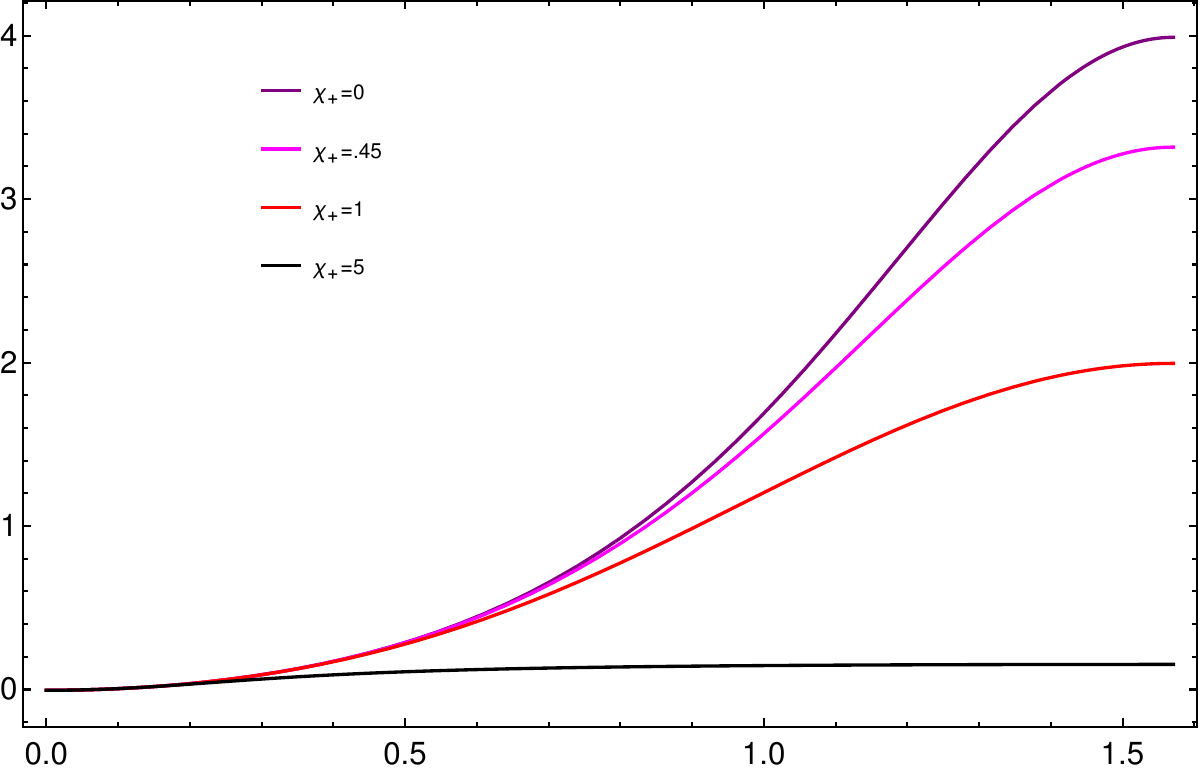}   
    \caption{Potential $V(\theta)$ plot for $\chi_{+}$=$\chi_{-}$$=\chi$ (left) and  for different values of $\chi_{+}$ fixing $\chi_{-}=1$,(right), for both the plots $m=2$ and $\mathcal{E}=0.1$}
    \label{fig 1}
\end{figure}

\begin{figure}[ht!]
    \centering
\includegraphics[width=0.45\textwidth,height=4.5cm]{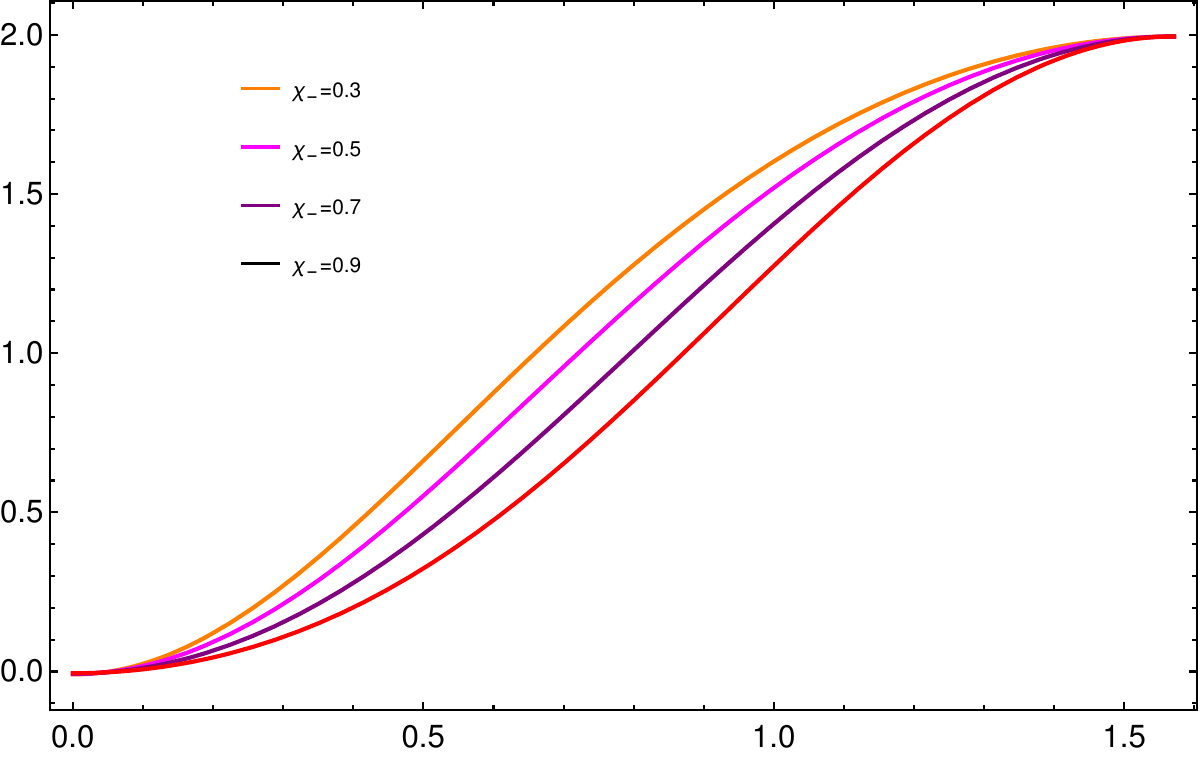}\hfill
\includegraphics[width=0.45\textwidth,height=4.5cm]{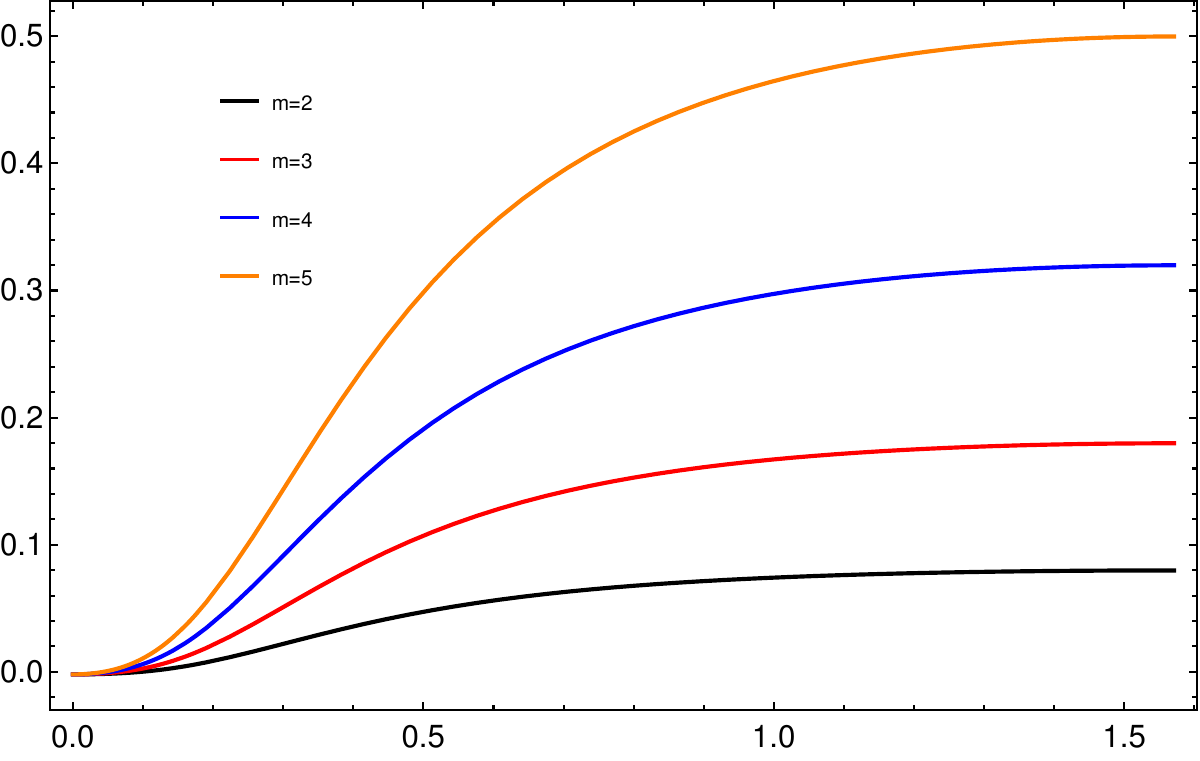}   
    \caption{Potential $V(\theta)$ plot for different values of $\chi_{-}$ at fixed $\chi_{+}=1$,$m=2$ and $\mathcal{E}=0.1$ (left) and for different values of $m$ (right) where the other  parameters are fixed at  $\chi_{+}=7$,$\chi_{-}=2$ and $\mathcal{E}=0.1$.}
    \label{fig 2}
\end{figure}

Equation (\ref{virasoro}) can also be written in terms of an effective potential $V(\theta)$ such as
$$\Pi_{\theta}^2+V(\theta)=0 ,$$
where 
 \begin{align}
     \Pi_{\theta}=\frac{\Dot{\theta}}{(1+\chi_-^2cos^2{\theta}+\chi_+^2sin^2{\theta})}
    \label{canon}, \\
    V(\theta)= \frac{m^2 \sin ^2\theta \left(1+\chi _{+}^2 \sin ^2\theta\right)}{\left(1+\chi _{+}^2 \sin ^2\theta+\chi _{-}^2 \cos ^2\theta\right)^2}-\frac{\mathcal{E}^2}{1+\chi _{+}^2 \sin ^2\theta+\chi _{-}^2 \cos^2\theta}.
   \label{pot}
 \end{align}
and $\Pi_{\theta}$ is the canonical momentum corresponding to coordinate $\theta$. In figure \ref{fig 1} and \ref{fig 2} we show the nature of the potential for different values of the deformation parameters.
\hspace{5cm}
 \newline
 In order to get a deeper understanding of the string dynamics we compute the string oscillation number and its relation to energy. The oscillation number associated with the periodic angular movement can be expressed as
 \begin{equation}
     \mathcal{N}=\frac{1}{2\pi}\oint d\theta ~\Pi_{\theta} .
     \label{oscln}
 \end{equation}
Now substituting equation (\ref{virasoro}) in equation (\ref{oscln}) and $\sin^2{\theta}=x$, the oscillation number can be expressed as following
    \begin{equation}
     \mathcal{N}=\frac{1}{\pi}\int_{0}^{{Z_2}} \frac{dx}{((1+\chi_-^2)+(\chi_+^2-\chi_-^2)x)} ~\sqrt{\frac{-\mathcal{E}^2((1+\chi_-^2)+(\chi_+^2-\chi_-^2)x)+m^2x(1+\chi_+^2x)}{x(x-1)}}.   
     \label{oscln2}
    \end{equation} 
The above integral can be readily evaluated considering a short string limit where $\mathcal{E} \rightarrow 0$ and we obtain the oscillation number expressed as a series in $\mathcal{E}$  
\begin{equation}
    \begin{split}
      \mathcal{N}&=\frac{\mathcal{E}^2}{2 m}-\Big[\left(3 \chi _+^2+1\right) \chi _-^2+\chi _+^2-1\Big]\frac{ \mathcal{E}^4}{16 m^3}+\Big[\left(35 \chi _+^4+22 \chi _+^2+3\right) \chi _-^4\\& +2   \left(11 \chi _+^4-6 \chi _+^2-1\right) \chi _-^2+3 \chi _+^4-2 \chi _+^2+11\Big]\frac{\mathcal{E}^6}{128 m^5}+O\left(\mathcal{E}^7\right).
    \end{split}
\end{equation}
Performing an inversion of the aforementioned series, we can deduce the string energy in terms of the oscillation number as follows,
\begin{equation}
    \begin{split}
      \mathcal{E}&=\sqrt{2m\mathcal{N}} \Bigg[ 1+\Big[\left(3 \chi _+^2+1\right) \chi _-^2+\chi _+^2-1\Big]\frac{ \mathcal{N}}{8 m}-\Big[\left(77 \chi _+^4+46 \chi _+^2+5\right) \chi _-^4\\&+\left(46 \chi _+^4-20 \chi _+^2+6\right) \chi _-^2+(5 \chi _+^4+6 \chi _+^2+37)\Big]\frac{\mathcal{N}^2}{128 m^2} +O\left(\mathcal{N}^{3}\right)\Bigg].
    \end{split}
\end{equation}
One can check for $\chi_-$  $\rightarrow$ $0$ and $\chi_{+}$ $\rightarrow$ $0$, the  dispersion relation matches with the undeformed case studied in \cite{Beccaria:2010zn}. 
The above result can also be verified by performing the quantisation starting with a Nambu-Goto action. 
\section{Perturbation of pulsating string in two parameter deformed $\mathcal{R} \times S^2$ }
In this section, we use the covariant approach developed by Vega and Sanchez \cite{deVega:1987veo} to study the perturbation of bosonic string moving in an arbitrary curved space-time. The equations governing the perturbations are the well-known Jacobi equations and are derived from the second variation of the action (\ref{polyakov}). The variation of the embedding coordinates can be expressed in terms of normal ($N_{i}^{\mu}$) and tangent vectors ($X^{\mu}_{,\alpha}$) as 
$\delta X^{\mu}=\Phi^{i}N_{i}^{\mu}+\psi^{\alpha}X^{\mu}_{,\alpha}$. Here $\Phi^{i}$ and $\psi^{\alpha}$ are the perturbations along normal and tangent directions respectively. Here $i$ denotes the number index. 
The normal vectors satisfy the orthogonality condition
\begin{equation}
g_{\mu\nu}N^{\mu}_{i}N^{\nu}_{j}=\delta_{ij}~~,~~g_{\mu\nu}N^{\mu}_{i}\partial_{\alpha}X^{\nu}=0.
\end{equation}
The extrinsic curvature and normal fundamental forms are defined by
\begin{equation}
    K_{i,\alpha\beta}=g_{\mu\nu}N^{\mu}_{i}X^{\rho}_{,\alpha}\nabla_{\rho}X^{\nu}_{,\beta} ,~~~\mu_{ij,\alpha}=g_{\mu\nu}N^{\mu}_{i}\partial_{\alpha}X^{\rho}\nabla_{\rho}N^{\nu}_{j},
\end{equation}
where $\nabla_{\rho}$ is the covariant derivative w.r.t space time coordinates. Now considering all this the first order perturbation equation satisfied by $\varphi^{i}$ is given by \cite{Larsen:1993mx},\cite{Khan:2005fc} 
\begin{equation}
    \Big(\delta^{kl} h^{\alpha\beta}D_{ik\alpha} D_{lj\beta}+\frac{2}{G^{c}_{c}} K^{\alpha\beta}_{i} K_{j,\alpha\beta} -h^{\alpha\beta}R_{\mu\rho\sigma\nu}N^{\rho}_{i}N^{\sigma}_{j}X^{\mu}_{,\alpha}X^{\nu}_{,\beta}\Big)\varphi^{j}=0,
\end{equation}
where $D_{ij\alpha}=\delta_{ij}D_{\alpha}+\mu_{ij\alpha}$.
 The background metric is the two parameter deformed $\mathcal{R} \times S^2$ given in section 2.
From equation (\ref{ansatz}) and equation (\ref{induced}) the induced metric is given by
\begin{eqnarray}
    G_{\alpha \beta} =\frac{m^2\sin^2{\theta}(1+\chi_+^2\sin^2{\theta})}{1+\chi_-^2cos^2{\theta}+\chi_+^2sin^2{\theta}}\left(-d\tau^2+d\sigma^2  \right) .
\end{eqnarray}
The tangent and normal vectors to the worldsheet are given by
\begin{equation}
    \dot{X}=(\mathcal{E},\dot{\theta},0)~,~~~X'=(0,0,m)
\end{equation}
\begin{eqnarray}
N^{\mu}=  \bigg \{ \frac{\sqrt{\Tilde{F}(\theta)\mathcal{E}^2-m^2\sin^2{\theta}(1+\chi_+^2\sin^2{\theta})}}{m\sin{\theta}\sqrt{1+\chi_{+}^2\sin^2{\theta}}}\, , \frac{\mathcal{E\Tilde{F}(\theta)}}{m\sin{\theta}\sqrt{1+\chi_{+}^2\sin^2{\theta}}}\, ,0\bigg\}. 
\end{eqnarray}
Here all the normal fundamental forms vanish. The extrinsic curvature tensor of the worldsheet is found to be 
\begin{eqnarray}
    &&  K_{\tau\tau}= -\frac{m \mathcal{E} \cos {\theta} \left(\left(\chi _+^2 \sin ^2{\theta}+1\right) \left(\left(\chi _-^2+\chi _+^2\right) \sin ^2{\theta}+1\right)+\chi _-^2 \cos ^2{\theta} \left(2 \chi _+^2 \sin ^2{\theta}+1\right)\right)}{\sqrt{\chi _+^2 \sin ^2{\theta}+1} \left(\chi _+^2 \sin ^2{\theta}+\chi _-^2 \cos ^2{\theta}+1\right)},\nonumber\\&&
    K_{\sigma\sigma}= -\frac{m \mathcal{E} \cos {\theta} \left(\left(\chi _+^2 \sin ^2{\theta}+1\right) \left(\left(\chi _-^2+\chi _+^2\right) \sin ^2{\theta}+1\right)+\chi _-^2 \cos ^2{\theta} \left(2 \chi _+^2 \sin ^2{\theta}+1\right)\right)}{\sqrt{\chi _+^2 \sin ^2{\theta}+1} \left(\chi _+^2 \sin ^2{\theta}+\chi _-^2 \cos ^2{\theta}+1\right)},\nonumber\\&&
K_{\tau\sigma}=K_{\sigma\tau}=0.
\end{eqnarray}
Using all the quantities mentioned above, we now proceed to derive the perturbation equation and obtain the following equation satisfied by the scalar $\varphi$
\begin{multline}
 \Phi''(\tau ,\sigma )-\Ddot{\Phi}(\tau
   ,\sigma )+\left[\frac{2 \left(\chi _-^2+1\right) \mathcal{E} ^2 }{\sin ^2{\theta}}+\frac{2 \chi _-^2 \left(2 \chi _+^4+5 \chi _+^2+3\right) \mathcal{E} ^2}{\chi_{+}^2(1+\chi _+^2 \sin^2{\theta})} \right. \\ \left. -\frac{3 \chi _-^2 \left(\chi _+^2+1\right)^2\mathcal{E} ^2}{\chi _+^2 \left(1+\chi_+^2\sin^2 {\theta}\right)^2} -\frac{\chi _-^2 \left(5 \chi _+^2+3\right) \mathcal{E}^2}{\chi _+^2}+(\chi _+^2-1)\mathcal{E}^2\right]\Phi(\tau ,\sigma)=0.
   \label{perteq}
\end{multline}
Let us take  a fourier expansion $\Phi(\tau, \sigma)=\sum_{n} \epsilon_0 \, \alpha_{n}(\tau)e^{in\sigma}$ under which equation (\ref{perteq}) reduces to 
\begin{eqnarray}
  \Ddot{\alpha_n}(\tau)+\Bigg[n^2-\mathcal{E}^2\left(\chi_{+} ^2-1\right)-\frac{2\mathcal{E}^2\left(\chi_{-} ^2+1\right)}{\sin^2{\theta}}-\frac{2 \chi _-^2 \left(2 \chi _+^4+5 \chi _+^2+3\right) \mathcal{E} ^2}{\chi_{+}^2(1+\chi _+^2 \sin^2{\theta})} \nonumber&&\\ +\frac{3 \chi _-^2 \left(\chi _+^2+1\right){}^2\mathcal{E} ^2}{\chi _+^2 \left(1+\chi_+^2\sin^2 {\theta}\right){}^2} +\frac{\chi _-^2 \left(5 \chi _+^2+3\right) \mathcal{E}^2}{\chi _+^2}\Bigg]\alpha_n(\tau)=0. 
  \label{pert2}
\end{eqnarray}
where $\epsilon_0 << 1$  can be considered as the amplitude of perturbation. One can rearrange the equation (\ref{pert2}) to  get a second order differential equation of the form 
\begin{equation}
    \frac{d^2Y}{d\tau^2}+\{n^2-V(\tau)\}Y=0,\label{perteq310}
\end{equation}
where $V(\tau)$ is given by,
\begin{eqnarray}
  V(\tau)&= &\mathcal{E}^2 \Bigg[(\chi _+^2-1)-\frac{\left(5 \chi _+^2+3\right) \chi _{-}^2}{\chi _+^2}+\frac{2 \left(\chi _-^2+1\right)}{z(\tau)^2}+\cfrac{2 \chi _-^2 \left(2 \chi _+^4+5 \chi _+^2+3\right)}{\chi _+^2 \left(1+z(\tau)^2\right)}\nonumber\\&& \hspace{8.5cm}-\cfrac{3 \chi _-^2 \left(\chi _+^2+1\right)}{\chi _+^2\left(1+z(\tau) ^2\right)^2}\Bigg]
 \end{eqnarray}

\begin{figure}[h!]
	\centering
\includegraphics[width=0.45\linewidth]{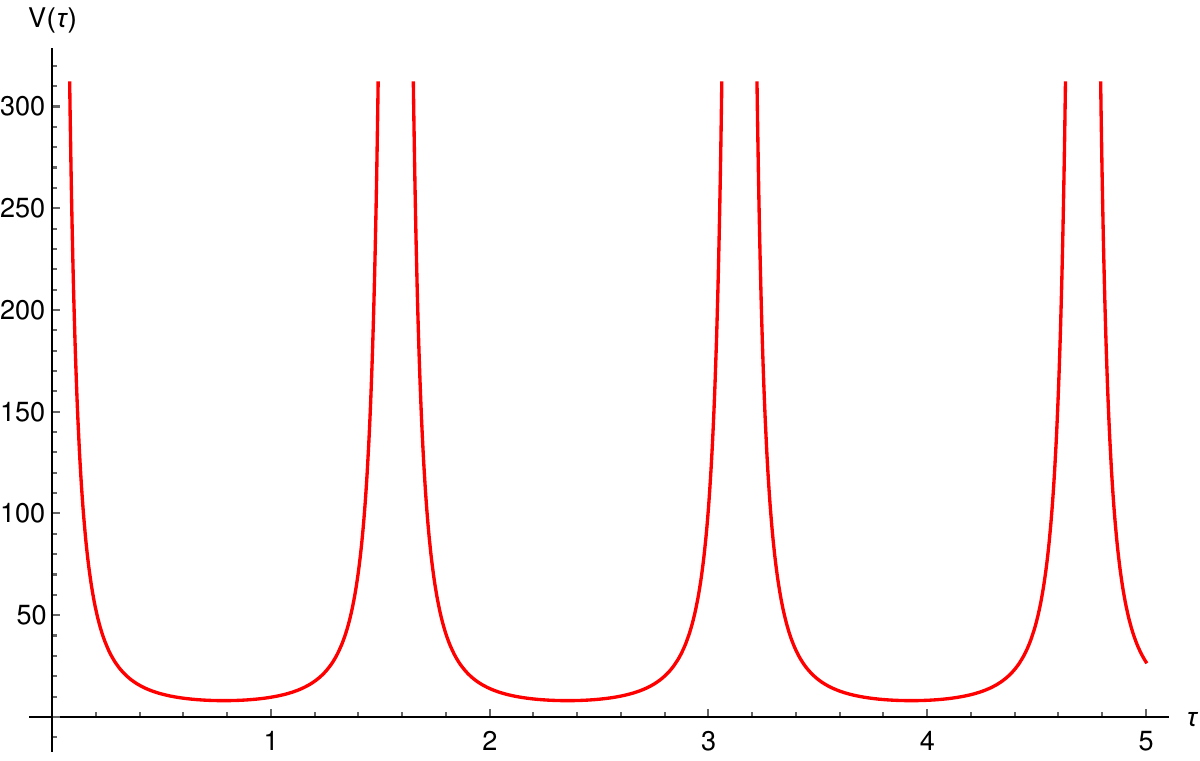}\hfill
\includegraphics[width=0.45\linewidth]{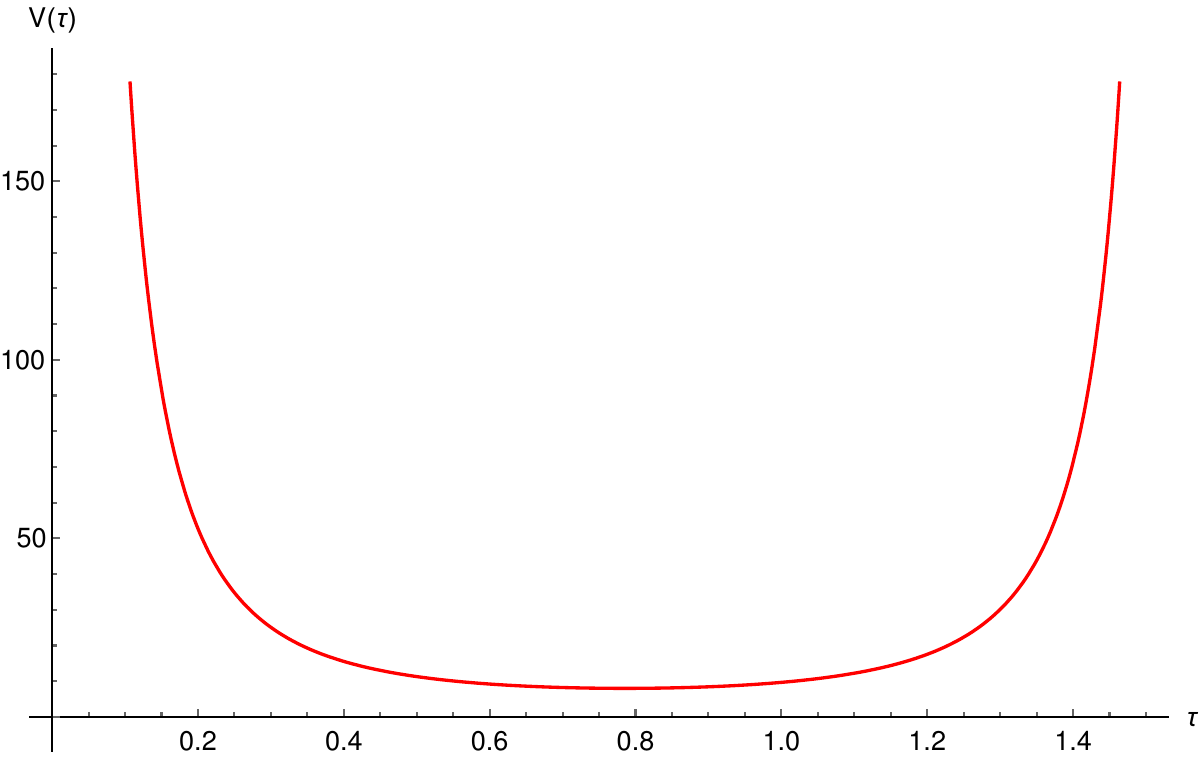}
    \caption{V($\tau$) vs  $\tau$ plot for $\mathcal{E}$=0.03, m=2, $\chi_{+}$=0.5, $\chi_{-}$=0.3 . }
    \label{fig:3}
\end{figure}

 Figure \ref{fig:3} shows the nature of potential $V(\tau)$ for $\mathcal{E}$=0.03, m=2, $\chi_{+}$=0.5, $\chi_{-}$=0.3.
The potential corresponding to the perturbation equation is a well-like potential and hence, there will be bound state solutions corresponding to the normal mode fluctuations. Now considering the complicated expression of $V(\tau)$, it is a difficult task to solve the second order differential  equation (\ref{perteq310}) analytically. We next try to simplify the complexity of the problem by setting one of the deformation parameter to zero.
\subsection{Perturbation of pulsating string $\mathcal{R} \times S_{\chi}^2$}
In this subsection, we want to analyse the perturbation of string in a simplified scenario where one of the deformation parameters from the two-parameter case is set to zero. By fixing one parameter to zero, we reduce the complexity of the background deformation, allowing us to do an analytic study on the perturbation equation. Now setting $\chi_{+}=0$ and $\chi_{-}=\chi$, the metric equation (\ref{2.2}) reduces to
\begin{equation}
    ds^2=-dt^2+\frac{d\theta^2}{1+\chi^2\cos^2{\theta}}+\frac{\sin^2{\theta}d\phi^2}{1+\chi^2\cos^2{\theta}}.
\end{equation}
Considering the pulsating string ansatz as given in equation (\ref{ansatz}) and from equation (\ref{virasoro}) the Virasoro equation reduces to 
\begin{equation}
    \dot{\theta}^2=\mathcal{E}^2(1+\chi^2\cos^2{\theta})-m^2\sin^2{\theta},
\end{equation}
and the solution to the theta equation is given by,
\begin{equation}
    \sin{\theta}=\sqrt{\frac{\mathcal{E}^2(1+\chi^2)}{m^2+\mathcal{E}^2\chi^2}} ~~\textbf{sn} \left(\sqrt{m^2+\mathcal{E}^2\chi^2} \tau ~|~\frac{\mathcal{E}^2(1+\chi^2)}{m^2+\mathcal{E}^2\chi^2} \right),
    \label{sin}
\end{equation}
Following a similar analysis as the previous section we get the reduced perturbation equation as
\begin{equation}
  \Phi''(\tau ,\sigma )-\Ddot{\Phi}(\tau ,\sigma )+  \mathcal{E}^2\left(\chi ^2+1\right)(1+2\cot^2{\theta})\Phi(\tau,\sigma)=0.
\label{diffeqn}
\end{equation}
%
To solve the perturbation equation analytically we first choose to consider a short string limit. In this limit neglecting the higher order terms in $\frac{\mathcal{E}}{m}$, equation (\ref{sin})  and equation (\ref{diffeqn}) correspondingly take the form, 
\begin{equation}
    \sin{\theta}=\frac{\mathcal{E}\sqrt{(1+\chi^2)}}{m}~~\sin{ m \tau},
\end{equation}
\begin{equation}
    \Ddot{\alpha_n}(\tau)+\Big( n^2-\frac{2m^2}{\sin^2{m\tau}}+\mathcal{E}^2\left(\chi ^2+1\right)\Big)\alpha_n(\tau)=0.
\label{oneparaeq}
\end{equation}
where we have used $\Phi(\tau, \sigma)=\sum_{n} \epsilon_0 \, \alpha_{n}(\tau)e^{in\sigma}$. From equation (\ref{oneparaeq}) it is clearly evident that the differential equation is a special case of the well-known Poschl-Teller equation and a general solution for such a differential equation is given in the form
\begin{equation}
    \alpha_n(\tau)=C_1 P_1(\tau)+C_2 P_2(\tau)
\end{equation}
where,
\begin{align}
    P_1(\tau)=\sin^2{m\tau} ~_2F_1\left(1+\frac{\sqrt{n^2+\mathcal{E}^2(1+\chi^2)}}{2m},1-\frac{\sqrt{n^2+\mathcal{E}^2(1+\chi^2)}}{2m},\frac{5}{2};\sin^2{m\tau} \right) \nonumber\\
P_2(\tau)=\frac{1}{\sin{m\tau}}  ~_2F_1     \left(\frac{\sqrt{n^2+\mathcal{E}^2(1+\chi^2)}-m}{2m},\frac{-\sqrt{n^2+\mathcal{E}^2(1+\chi^2)}-m}{2m},-\frac{1}{2};\sin^2{m\tau} \right)
\end{align}

and $C_1$,$C_2$ are constants.

\begin{figure}[h]
\includegraphics[scale=0.39]{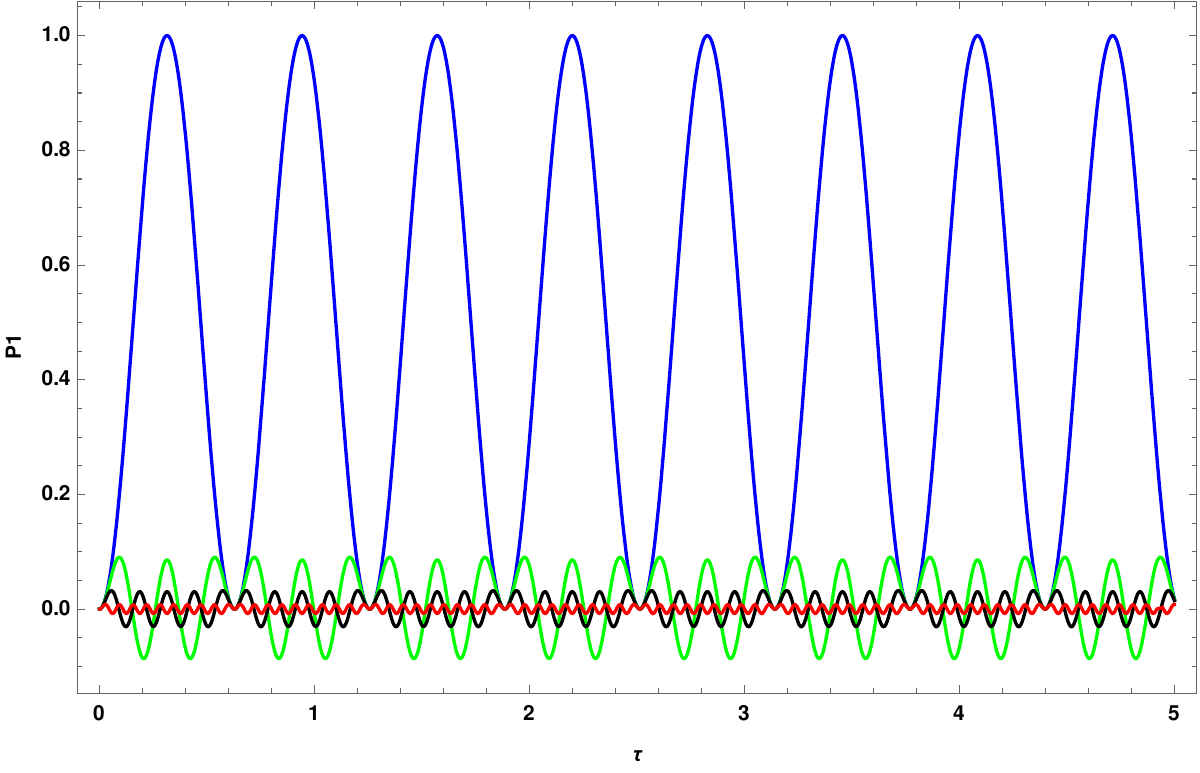}
\includegraphics[scale=0.39]{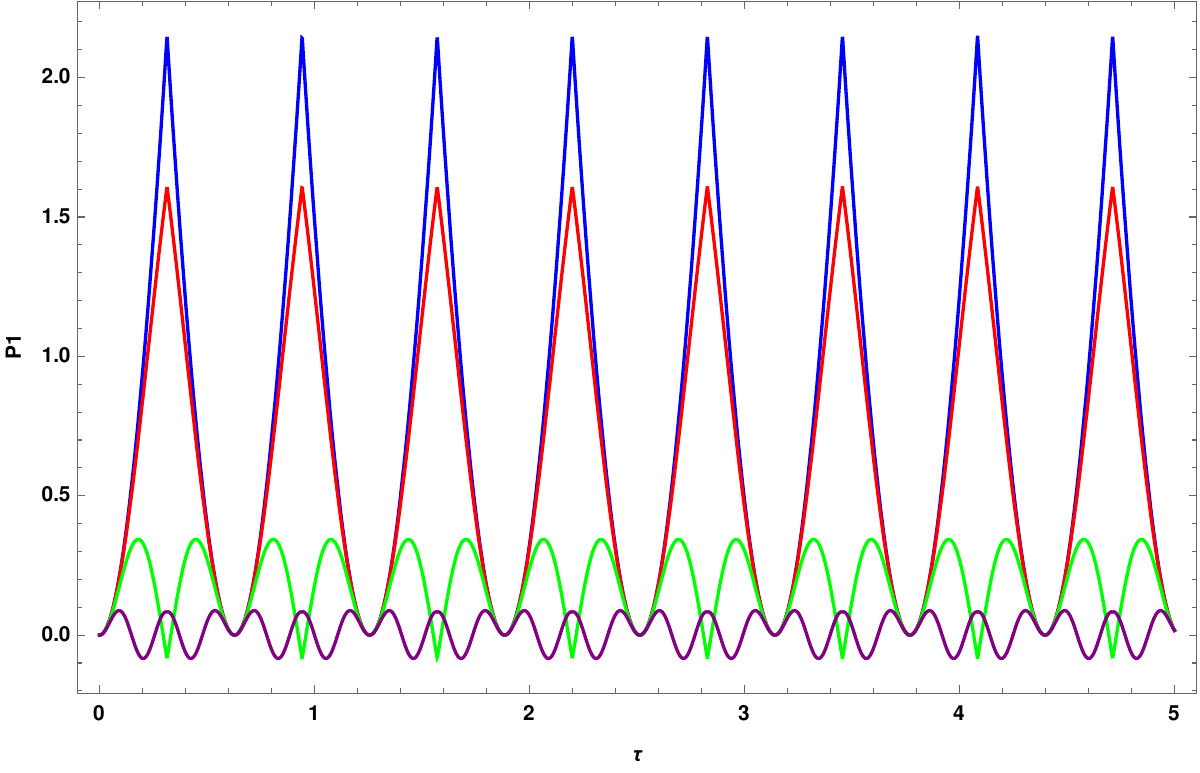}
\caption{(a)Plot showing nature of solution $P_1(\tau)$ for for different values of  $n=$(10(blue),30(green),50(black),100(red)), $m=5$,~$\mathcal{E}=0.03,~\chi=0.09$, and (b)behavior of $P_1(\tau)$ for for different values of $\chi =$(100(blue),200(red),500(green),1000(purple)) ,$m=5$,~$\mathcal{E}=0.03$,~$n=5$. From both the plots we observe that the amplitude of oscillations decreases for higher values of $n$ and $\chi$.  }
\label{fig6}
\end{figure}

\begin{figure}[h!]
\includegraphics[scale=0.39]{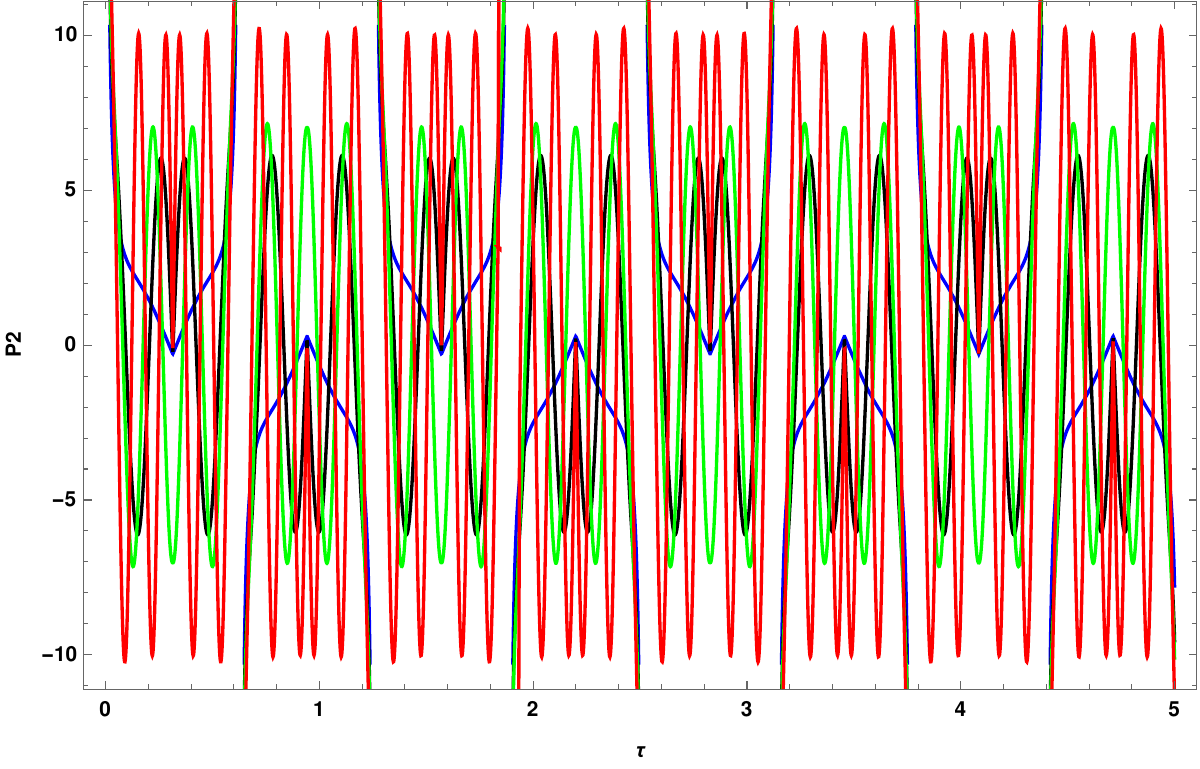}
\includegraphics[scale=0.39]{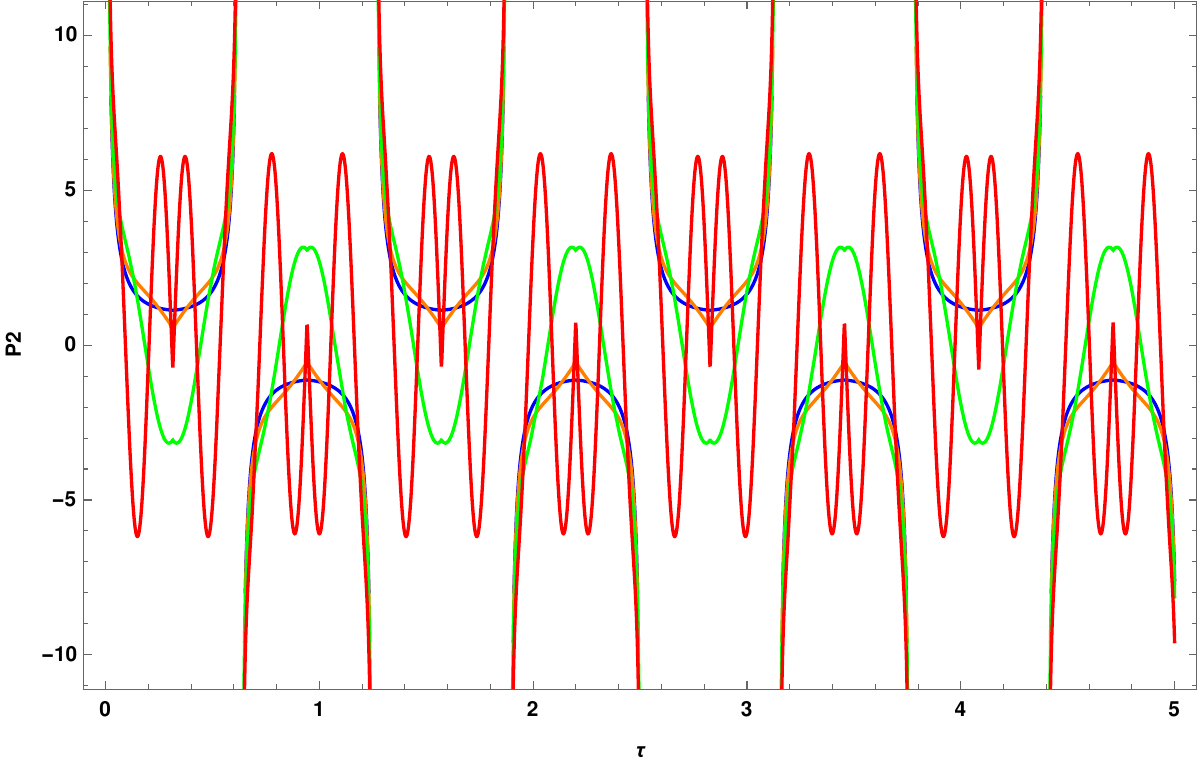}
\caption{(a)Behavior of solution $P_2(\tau)$ for different values of $n=$ (10(blue),30(black),35(green),50(red)) and m=5,~$\mathcal{E}=0.03,~\chi=100$,(b)Behavior of solution $P_2(\tau)$ for different values of  $\chi=$ (100(blue),250(orange),500(green),1000(red))  and m=5,~$\mathcal{E}=0.03$,~$n=5$.}
\label{fig7}
\end{figure}

Now the scalar function can be written as 
\begin{equation}
    \Phi(\tau,\sigma)=\sum_{n} \epsilon_{0}\left(C_1 P_1(\tau)+C_2 P_2(\tau)\right)e^{in\sigma }
\end{equation} 
Now to understand the nature of the solution to the perturbation equation we have plotted $P_1(\tau)$ and $P_2(\tau)$. Figure (\ref{fig6}) shows $P_1(\tau)$ for different values of $n$ (left) and $\chi$ (right). It is observed that the solution $P_1(\tau)$ is oscillatory and the amplitude of perturbation decreases with increasing values of both $n$ and $\chi$. Similarly, the solution $P_2(\tau)$ is shown in figure (\ref{fig7}). Unlike the first solution, it is diverging periodically. Hence, $P_2(\tau)$ does not contribute to the physical perturbations.   
Now, the physical perturbations for short string solution can be written as
\begin{equation}
    \delta t=N^{t}\Phi(\tau ,\sigma)=\epsilon_0 \sum_{n}\sin{m\tau}\sqrt{1-\left(1+\frac{\mathcal{E}^2\chi^2}{m^2}\right)\sin^2{m\tau}} ~P_1(\tau)\cos{n\sigma}
\end{equation}
\begin{equation}
    \delta \theta=N^{\theta}\Phi(\tau ,\sigma)=\epsilon_0 \sum_{n}\sin{m\tau}\frac{1+\chi^2\left(1-\frac{\mathcal{E}^2(1+\chi^2)}{m^2}\sin^2{m\tau}\right)}{\sqrt{1+\chi^2}} ~ P_1(\tau)\cos{n\sigma}
\end{equation}
\begin{equation}
    \delta \phi=N^{\phi}\Phi(\tau ,\sigma)=0
\end{equation}

\medskip
Let's discuss the perturbation without considering the short string limit. Then substituting equation (\ref{sin}) in the perturbation equation (\ref{diffeqn}) and using $\Phi(\tau, \sigma)=\sum_{n} \alpha_{n}(\tau)e^{in\sigma}$, we get 
\begin{equation}
    \Ddot{\alpha_n}(\tau)+\Bigg( n^2-\cfrac{2(m^2+\mathcal{E}^2\chi^2)}{sn^2\left(\sqrt{m^2+\mathcal{E}^2\chi^2} \tau~ |~\frac{\mathcal{E}^2(1+\chi^2)}{m^2+\mathcal{E}^2\chi^2} \right)}+\mathcal{E}^2\left(\chi ^2+1\right)\Bigg)\alpha_n(\tau)=0.
    \label{difeq1}
\end{equation}
Let's take $sn^2\left(\sqrt{m^2+\mathcal{E}^2\chi^2} \tau |\frac{\mathcal{E}^2(1+\chi^2)}{m^2+\mathcal{E}^2\chi^2} \right))=u$ in equation (\ref{difeq1}) and with a substitution $\alpha_n=us$ we have,
\begin{eqnarray}
 \frac{d^2s}{du^2}+\frac{1}{2}\Bigg[ \frac{5}{u}+\frac{1}{u-1}+\cfrac{1}{u-\cfrac{\mathcal{E}^2\chi^2+m^2}{\mathcal{E}^2(1+\chi^2)}}~~\Bigg]\cfrac{ds}{du}+\frac{\Bigg[\frac{3}{2}u+\frac{n^2-4(\mathcal{E}^2\chi^2+m^2))}{4\mathcal{E}^2(1+\chi^2)}-\frac{3}{4} \Bigg]}{u(u-1)\bigg(u-\cfrac{\mathcal{E}^2\chi^2+m^2}{\mathcal{E}^2(1+\chi^2)}\bigg)}s=0 .
\end{eqnarray}
This is exactly the canonical form of Heun's general equation and the solution to the above differential equation can be written as in terms of Heun's functions.
$$s=H\bigg(\cfrac{\mathcal{E}^2\chi^2+m^2}{\mathcal{E}^2(1+\chi^2)},\frac{3}{4}-\frac{n^2-4(\mathcal{E}^2\chi^2+m^2))}{4\mathcal{E}^2(1+\chi^2)};1,\frac{3}{2},\frac{5}{2},1;sn^2\left(\sqrt{m^2+\mathcal{E}^2\chi^2} \tau ~|~\frac{\mathcal{E}^2(1+\chi^2)}{m^2+\mathcal{E}^2\chi^2} \right)\bigg)$$
and \begin{eqnarray}
    &&\alpha_n(\tau)=sn^2\left(\sqrt{m^2+\mathcal{E}^2\chi^2} \tau ~~|~~\frac{\mathcal{E}^2(1+\chi^2)}{m^2+\mathcal{E}^2\chi^2} \right)\nonumber\\&&H\bigg(\cfrac{\mathcal{E}^2\chi^2+m^2}{\mathcal{E}^2(1+\chi^2)},\frac{3}{4}-\frac{n^2-4(\mathcal{E}^2\chi^2+m^2))}{4\mathcal{E}^2(1+\chi^2)};1,\frac{3}{2},\frac{5}{2},1;sn^2\left(\sqrt{m^2+\mathcal{E}^2\chi^2} \tau~~ |~~\frac{\mathcal{E}^2(1+\chi^2)}{m^2+\mathcal{E}^2\chi^2} \right)\bigg)\nonumber
\end{eqnarray}
\subsection{Perturbation of pulsating string $AdS_{3_{\chi}}$}
In this subsection, we study the worldsheet perturbation of the pulsating string in the deformed $AdS_3$ background where $\chi_{+}=0$, $\chi_{-}=\chi$ and the spherical coordinates are not in use. We begin by considering the metric
\begin{equation}
    ds^2=-\frac{\cosh^2{\rho}}{1-\chi^2 \sinh^2{\rho}}dt^2+\frac{d\rho^2}{1-\chi^2\sinh^2{\rho}}+\sinh^2{\rho}d\phi^2.
\end{equation}

To avoid the singularity surface present in this coordinate system as explained in \cite{Banerjee:2016xbb},\cite{Kameyama:2014vma} we use a coordinate transformation following \cite{Kameyama:2014vma}
$$\cosh{\Theta}=\frac{\cosh{\rho}}{\sqrt{1-\Theta^2\sinh^2{\rho}}},$$
under which the metric transforms to 
\begin{eqnarray}
    ds^2=-\cosh^2{\Theta}dt^2+\frac{d\Theta^2}{1+\chi^2\cosh^2{\Theta}}+\frac{\sinh^2{\Theta}}{1+\chi^2\cosh^2{\Theta}}d\phi^2.
\end{eqnarray}
Now considering  a pulsating string ansatz,
 \begin{equation}
     t=t(\tau),~\Theta=\Theta(\tau),~\phi=m\sigma,
 \end{equation}
we obtain the equation of motion for coordinate $t$ and $\Theta$ as follows,
\begin{align}
    \dot{t}=\frac{\mathcal{E}}{\cosh^2{\Theta}},\\
\Dot{\Theta}^2=\frac{\mathcal{E}^2(1+\chi^2 \cosh^2{\Theta})}{\cosh^2{\Theta}}-m^2\sinh^2{\Theta}.
\label{Thetaeq}
\end{align}
In this new coordinate system solving the above differential equation (\ref{Thetaeq}), we get the solution
\begin{equation}
    \sinh{\Theta}=\sqrt{\frac{-R_{-}R_{+}}{R_{+}-R_{-}}}\textbf{sd}\Big(m\sqrt{R_{+}-R_{-}} \tau~~|~~\frac{R_{+}}{R_{+}-R_{-}}\Big),
    \label{sinh}
\end{equation}
where,$$R_{\pm}=\frac{(\mathcal{E}^2\chi^2-m^2)\pm \sqrt{(\mathcal{E}^2\chi^2-m^2)^2+4m^2\mathcal{E}^2(1+\chi^2)}}{2m^2}.$$
For more details on the solution one can check \cite{Banerjee:2016xbb}.
Now, the induced metric on the worldsheet is given by
\begin{equation}
    ds^2=\frac{m^2\sinh^2{\Theta}}{1+\chi^2\cosh^2{\Theta}}\left( -d\tau^2+d\sigma^2\right).
\end{equation}
The tangent vectors are given by 
\begin{equation}
    \Dot{X}^{\mu}=\left( \frac{\mathcal{E}}{\cosh^2{\Theta}} ,~~~~ \Dot{\Theta}, ~~~0\right) ,~~~~~~X'^{\mu}=\left(0,~~ 0, ~~m \right) ,
\end{equation}
and  normal vector is given by
\begin{eqnarray}
   N^{\mu}=&&\Bigg(\frac{\sqrt{\mathcal{E} ^2 \left(\text{sech}^2{\Theta
   }+\chi ^2\right)-m^2 \sinh ^2(\Theta)}}{m\sinh{\Theta}\cosh{\Theta}}~, ~\frac{\mathcal{E}   \left(\chi ^2 \cosh ^2{\Theta}+1\right)}{m\sinh{\Theta}\cosh{\Theta}}~,~0\Bigg)  . 
\end{eqnarray}
 Now the components of the extrinsic curvature are given by
\begin{equation}
    K_{\tau\tau}= -\frac{m\mathcal{E} \left(\chi ^2+1\right) }{1+\chi^2\cosh^2{\Theta}}, \quad K_{\tau\sigma}=0, \quad K_{\sigma\tau}=0, \quad K_{\sigma\sigma}=-\frac{m\mathcal{E} \left(\chi ^2+1\right) }{1+\chi^2\cosh^2{\Theta}}~.
\end{equation}
Here all the normal fundamental forms vanishes. Hence the first order perturbation equation takes the form
\begin{equation}
  \Phi''(\tau ,\sigma )-\Ddot{\Phi}(\tau
   ,\sigma )+ \left(m^2-m^2\cosh{2 \Theta }+2  \mathcal{E} ^2 (1+\chi^2)\csch^2{\Theta}\right)\Phi(\tau,\sigma)=0.
\end{equation}
Now taking a Fourier expansion, $\Phi(\tau, \sigma)=\sum_{n} \epsilon_0 \alpha_{n}(\tau)e^{in\sigma}$, the above differential equation takes the form
\begin{equation}
  \Ddot{\alpha}_n(\tau)+\Big(n^2+2m^2\sinh^2{ \Theta}-\frac{2\mathcal{E}^2(1+\chi^2)}{\sinh^2{\Theta}}\Big)\alpha_n(\tau)=0.
\end{equation}
Now substituting the $\sinh{\Theta}$ value from equation  (\ref{sinh}) , we have 
\begin{align}
   && \Ddot{\alpha}_n(\tau)+\Bigg[n^2-2m^2\frac{R_{-}R_{+}}{R_{+}-R_{-}}\textbf{sd}^2\Big(m\sqrt{R_{+}-R_{-}} ~~|~~\frac{R_{+}}{R_{+}-R_{-}}\Big)\nonumber\\&&+\frac{2\mathcal{E}^2(1+\chi^2)}{\frac{R_{-}R_{+}}{R_{+}-R_{-}}\textbf{sd}^2\Big(m\sqrt{R_{+}-R_{-}} ~~|~~\frac{R_{+}}{R_{+}-R_{-}}\Big)}\Bigg]\alpha_n(\tau)=0.
    \label{alphan}
\end{align}
For a short string limit  neglecting the higher order terms in $\frac{\mathcal{E}}{m}$, we have equation (\ref{alphan}) reduces to 
\begin{equation}
  \Ddot{\alpha}_n(\tau)+\Big(n^2-\frac{2m^2}{\sin^2{m\tau}}\Big)\alpha_n(\tau)=0.
  \label{alpha2}
\end{equation}
Let us  consider $y=\sin^2{m\tau}$. Equation (\ref{alpha2}) takes the form
$$y(1-y)\alpha_n^{''}+(\frac{1}{2}-y)\alpha_n^{'}+\frac{1}{4}\left(\frac{n^2}{m^2}-\frac{2}{y}\right)\alpha_n=0.$$
From the above equation, it is clear that the perturbation equation in the deformed background in a short string limit is independent of the deformation parameter. The solution to the above differential equation has a similar form as the undeformed case.
\begin{eqnarray}
    \alpha_n(\tau)=C_1 \sin^2{m \tau } P_1(\tau) +\frac{C_2 } {\sin{m \tau}}P_2(\tau)
\end{eqnarray}
\begin{align}
    P_1(\tau)=\, _2F_1\left(1-\frac{n}{2 m},\frac{n}{2 m}+1;\frac{5}{2};\sin ^2(m \tau )\right)\\
    P_2(\tau)=i \, _2F_1\left(-\frac{n}{2 m}-\frac{1}{2},\frac{n}{2 m}-\frac{1}{2};-\frac{1}{2};\sin ^2(m \tau )\right)
\end{align}
 \\
 Analysing the behavior of these two solutions $P_1(\tau)$ and $P_2(\tau)$ for a short string limit, we conclude that one of the solutions $P_1(\tau)$  is oscillatory and hence defines the physical stable perturbation, whereas the other solution is divergent and hence is not stable. 
\section{Summary and Conclusion}
Let us first summarise the results obtained in this article. Firstly, we have considered the motion of a pulsating string in the novel two parameter deformed $\mathcal{R}\times S^2$ background. With the help of the Virasoro constraint, we found the string motion is equivalent to the motion of a particle under the influence of a potential. The nature of the potential is studied. We have derived the conserved quantities and shown a relation between the string energy and it's oscillation number in order to get a better understanding in the string dynamics. Next, we write down the fundamental equations describing the fluctuations of a fundamental string with pulsating nature in the two parameter deformed $\mathcal{R}\times S^2$ background. Because of the complicated form of the perturbation equation we have not solved the perturbation equation. We have simplified our analysis by considering one of the deformation parameters to be zero and study the perturbation equation in both $\chi$- deformed $\mathcal{R}\times S^2$ and $AdS_3$ background. Finally, we comment on the stability nature of the string solution by looking at the solutions governing perturbations. By examining the solutions in both an undeformed and a single-parameter deformed background, we anticipate similar behavior of the solutions in a two-parameter deformed background, which may include both stable and unstable solutions.
\medskip

We have shown our result is a generalization of the solutions obtained in the one parameter deformed background.
It would be an interesting direction to verify our result by reducing the current string sigma model into the well-known 1D integrable Neumann-Rosochatius model using a pulsating string ansatz. Exploring the dual gauge theory of such string background can be quite fascinating for future work. An extension to our work can be finding the dispersion relation in a three parameter deformed $\mathcal{R} \times S^2$ and  background constructed in \cite{Garcia:2021iox} in the presence of a mixed NS-NS and RR- field for any generic string solution.


\begin{thebibliography}{99}

\normalsize

\bibitem{Maldacena:1997re}
J.~M.~Maldacena,
``The Large N limit of superconformal field theories and supergravity,''
Adv. Theor. Math. Phys. \textbf{2}, 231-252 (1998)
doi:10.4310/ATMP.1998.v2.n2.a1
[arXiv:hep-th/9711200 [hep-th]].

\bibitem{Berenstein:2002jq}
D.~E.~Berenstein, J.~M.~Maldacena and H.~S.~Nastase,
``Strings in flat space and pp waves from N=4 superYang-Mills,''
JHEP \textbf{04}, 013 (2002)
doi:10.1088/1126-6708/2002/04/013
[arXiv:hep-th/0202021 [hep-th]].

\bibitem{Santambrogio:2002sb}
A.~Santambrogio and D.~Zanon,
``Exact anomalous dimensions of N=4 Yang-Mills operators with large R charge,''
Phys. Lett. B \textbf{545}, 425-429 (2002)
doi:10.1016/S0370-2693(02)02627-8
[arXiv:hep-th/0206079 [hep-th]].

\bibitem{Gubser:2002tv}
S.~S.~Gubser, I.~R.~Klebanov and A.~M.~Polyakov,
``A Semiclassical limit of the gauge / string correspondence,''
Nucl. Phys. B \textbf{636} (2002), 99-114
doi:10.1016/S0550-3213(02)00373-5
[arXiv:hep-th/0204051 [hep-th]].

\bibitem{Hofman:2006xt}
D.~M.~Hofman and J.~M.~Maldacena,
``Giant Magnons,''
J. Phys. A \textbf{39}, 13095-13118 (2006)
doi:10.1088/0305-4470/39/41/S17
[arXiv:hep-th/0604135 [hep-th]].

\bibitem{Kruczenski:2004wg}
M.~Kruczenski,
``Spiky strings and single trace operators in gauge theories,''
JHEP \textbf{08} (2005), 014
doi:10.1088/1126-6708/2005/08/014
[arXiv:hep-th/0410226 [hep-th]].

 \bibitem{Minahan:2002rc}
J.~A. Minahan, ``{Circular semiclassical string solutions on AdS(5) x S(5)},''
  {\em Nucl. Phys. B}, vol.~648, pp.~203--214, 2003.

\bibitem{Khan:2003sm}
A.~Khan and A.~L.~Larsen,
``Spinning pulsating string solitons in AdS(5) x S**5,''
Phys. Rev. D \textbf{69}, 026001 (2004)
doi:10.1103/PhysRevD.69.026001
[arXiv:hep-th/0310019 [hep-th]].

\bibitem{Dimov:2004xi}
H.~Dimov and R.~C.~Rashkov,
``Generalized pulsating strings,''
JHEP \textbf{05}, 068 (2004)
doi:10.1088/1126-6708/2004/05/068
[arXiv:hep-th/0404012 [hep-th]].

\bibitem{Smedback:2004udl}
M.~Smedback,
``Pulsating strings on AdS(5) x S**5,''
JHEP \textbf{07}, 004 (2004)
doi:10.1088/1126-6708/2004/07/004
[arXiv:hep-th/0405102 [hep-th]].

\bibitem{Kruczenski:2004cn}
M.~Kruczenski and A.~A.~Tseytlin,
``Semiclassical relativistic strings in S**5 and long coherent operators in N=4 SYM theory,''
JHEP \textbf{09}, 038 (2004)
doi:10.1088/1126-6708/2004/09/038
[arXiv:hep-th/0406189 [hep-th]].

\bibitem{Chen:2008qq}
B.~Chen and J.-B. Wu, ``{Semi-classical strings in AdS(4) x CP**3},'' {\em
  JHEP}, vol.~09, p.~096, 2008.

\bibitem{Dimov:2009rd}
H.~Dimov and R.~C. Rashkov, ``{On the pulsating strings in AdS(4) x CP**3},''
  {\em Adv. High Energy Phys.}, vol.~2009, p.~953987, 2009.

\bibitem{Bobev:2004id}
N.~P.~Bobev, H.~Dimov and R.~C.~Rashkov,
``Pulsating strings in warped AdS(6) x S**4 geometry,''
[arXiv:hep-th/0410262 [hep-th]].

\bibitem{Arnaudov:2010by}
D.~Arnaudov, H.~Dimov and R.~C.~Rashkov,
``On the pulsating strings in $AdS_5 x T^{1,1}$,''
J. Phys. A \textbf{44}, 495401 (2011)
doi:10.1088/1751-8113/44/49/495401
[arXiv:1006.1539 [hep-th]].

\bibitem{Park:2005kt}
I.~Y.~Park, A.~Tirziu and A.~A.~Tseytlin,
``Semiclassical circular strings in AdS(5) and 'long' gauge field strength operators,''
Phys. Rev. D \textbf{71}, 126008 (2005)
doi:10.1103/PhysRevD.71.126008
[arXiv:hep-th/0505130 [hep-th]].

\bibitem{Pradhan:2013sja}
P.~M.~Pradhan and K.~L.~Panigrahi,
``Pulsating Strings With Angular Momenta,''
Phys. Rev. D \textbf{88}, no.8, 086005 (2013)
doi:10.1103/PhysRevD.88.086005
[arXiv:1306.0457 [hep-th]].

\bibitem{Beccaria:2010zn}
M.~Beccaria, G.~V.~Dunne, G.~Macorini, A.~Tirziu and A.~A.~Tseytlin,
``Exact computation of one-loop correction to energy of pulsating strings in $AdS_5 x S^5$,''
J. Phys. A \textbf{44}, 015404 (2011)
doi:10.1088/1751-8113/44/1/015404
[arXiv:1009.2318 [hep-th]].

\bibitem{Giardino:2011jy}
S.~Giardino and V.~O.~Rivelles,
``Pulsating Strings in Lunin-Maldacena Backgrounds,''
JHEP \textbf{07}, 057 (2011)
doi:10.1007/JHEP07(2011)057
[arXiv:1105.1353 [hep-th]].

\bibitem{Banerjee:2014bca}
A.~Banerjee and K.~L.~Panigrahi,
``On the rotating and oscillating strings in (AdS$_{3}$  x S$^{3}$)$_{\kappa}$,''
JHEP \textbf{09}, 048 (2014)
doi:10.1007/JHEP09(2014)048
[arXiv:1406.3642 [hep-th]].

\bibitem{Panigrahi:2014sia}
K.~L.~Panigrahi, P.~M.~Pradhan and M.~Samal,
``Pulsating strings on (AdS$_{3}$ \texttimes{} S$^{3}$)$_{?}$,''
JHEP \textbf{03}, 010 (2015)
doi:10.1007/JHEP03(2015)010
[arXiv:1412.6936 [hep-th]].


\bibitem{Rotating2}
A.~Banerjee, K.~L.~Panigrahi and M.~Samal, 
{\em A note on oscillating strings in AdS$_{3}$ x S$^{3}$ with mixed three-form fluxes}, 
JHEP {\bf 1511} (2015) 133,
{\tt [arXiv:1508.03430 [hep-th]]}.


\bibitem{Barik1} 
  S.~P.~Barik, K.~L.~Panigrahi and M.~Samal,
 ``Perturbations of Pulsating Strings,''
  arXiv:1708.05202 [hep-th].

\bibitem{Banerjee:2014rza}
A.~Banerjee, S.~Biswas and K.~L.~Panigrahi,
``Semiclassical Strings in Supergravity PFT,''
Eur. Phys. J. C \textbf{74} (2014) no.10, 3115
doi:10.1140/epjc/s10052-014-3115-9
[arXiv:1403.7358 [hep-th]].


\bibitem{HN3} 
  R.~Hern\'andez, J.~M.~Nieto and R.~Ruiz,
  ``Pulsating strings with mixed three-form flux,''
  JHEP {\bf 1804}, 078 (2018)
  doi:10.1007/JHEP04(2018)078
  [arXiv:1803.03078 [hep-th]].

\bibitem{Banerjee:2016xbb}
A.~Banerjee and K.~L.~Panigrahi,
``On circular strings in $(AdS_3 \times S^3)_{\varkappa}$,''
JHEP \textbf{09} (2016), 061 doi: 10.1007/JHEP09 (2016)061
[arXiv:1607.04208 [hep-th]].

\bibitem{Biswas:2023uuq}
S.~Biswas, N.~Padhi and K.~L.~Panigrahi,
``Pulsating strings in I-brane background,''
Eur. Phys. J. C \textbf{84}, no.2, 208 (2024)
doi:10.1140/epjc/s10052-024-12543-w
[arXiv:2307.10765 [hep-th]].
\bibitem{Chakraborty:2022eeq}
A.~Chakraborty, N.~Padhi, P.~Pandit and K.~L.~Panigrahi,
``Neumann-Rosochatius system for strings on I-brane,''
JHEP \textbf{12}, 022 (2022)
doi:10.1007/JHEP12(2022)022
[arXiv:2209.09933 [hep-th]].

\bibitem{Larsen:1993iva}
A.~L.~Larsen and V.~P.~Frolov,
``Propagation of perturbations along strings,''
Nucl. Phys. B \textbf{414} (1994), 129-146
doi:10.1016/0550-3213(94)90425-1
[arXiv:hep-th/9303001 [hep-th]].

\bibitem{Lousto:1993vj}
C.~O.~Lousto and N.~G.~Sanchez,
``String instabilities in black hole space-times,''
Phys. Rev. D \textbf{47}, 4498-4509 (1993)
doi:10.1103/PhysRevD.47.4498
[arXiv:gr-qc/9212016 [gr-qc]].



\bibitem{Bhattacharya:2016ixc}
S.~Bhattacharya, S.~Kar and K.~L.~Panigrahi,
``Perturbations of spiky strings in flat spacetimes,''
JHEP \textbf{01} (2017), 116
doi:10.1007/JHEP01(2017)116
[arXiv:1610.09180 [hep-th]].
\bibitem{Bhattacharya:2018unr}
S.~Bhattacharya, S.~Kar and K.~L.~Panigrahi,
``Perturbations of spiky strings in AdS$_{3}$,''
JHEP \textbf{06} (2018), 089
doi:10.1007/JHEP06(2018)089
[arXiv:1804.07544 [hep-th]].

\bibitem{Bhattacharya:2021xfc}
S.~Bhattacharya, S.~Kar and K.~L.~Panigrahi,
``Perturbations of giant magnons and single spikes in $\mathbb R \times S^2$,''
[arXiv:2108.08622 [hep-th]].
\bibitem{Larsen:1994jt}
A.~L.~Larsen,
``Stable and unstable circular strings in inflationary universes,''
Phys. Rev. D \textbf{51} (1995), 4330-4336
doi:10.1103/PhysRevD.51.4330
[arXiv:hep-th/9403193 [hep-th]].
\bibitem{Kar:1997zi}
S.~Kar and S.~Mahapatra,
``Planetoid strings: Solutions and perturbations,''
Class. Quant. Grav. \textbf{15} (1998), 1421-1436
doi:10.1088/0264-9381/15/6/002
[arXiv:hep-th/9701173 [hep-th]].
\bibitem{Garriga:1991ts}
J.~Garriga and A.~Vilenkin,
``Perturbations on domain walls and strings: A Covariant theory,''
Phys. Rev. D \textbf{44} (1991), 1007-1014
doi:10.1103/PhysRevD.44.1007

\bibitem{Guven:1993ew}
J.~Guven,
``Covariant perturbations of domain walls in curved space-time,''
Phys. Rev. D \textbf{48} (1993), 4604-4608
doi:10.1103/PhysRevD.48.4604
[arXiv:gr-qc/9304032 [gr-qc]].

\bibitem{Guven:1993ex}
J.~Guven,
``Perturbations of a topological defect as a theory of coupled scalar fields in curved space,''
Phys. Rev. D \textbf{48}, 5562-5569 (1993)
doi:10.1103/PhysRevD.48.5562
[arXiv:gr-qc/9304033 [gr-qc]].

\bibitem{Larsen:1994ah}
A.~L.~Larsen and N.~G.~Sanchez,
``Strings propagating in the (2+1)-dimensional black hole anti-de Sitter space-time,''
Phys. Rev. D \textbf{50} (1994), 7493-7518
doi:10.1103/PhysRevD.50.7493
[arXiv:hep-th/9405026 [hep-th]].


\bibitem{Larsen:1993mx}
A.~L.~Larsen,
``Circular string instabilities in curved space-time,''
Phys. Rev. D \textbf{50}, 2623-2630 (1994)
doi:10.1103/PhysRevD.50.2623
[arXiv:hep-th/9311085 [hep-th]].
\bibitem{Khan:2005fc}
A.~Khan and A.~L.~Larsen,
``Improved stability for pulsating multi-spin string solitons,''
Int. J. Mod. Phys. A \textbf{21}, 133-150 (2006)
doi:10.1142/S0217751X06024888
[arXiv:hep-th/0502063 [hep-th]].
\bibitem{Larsen:2003ma}
A.~L.~Larsen and M.~A.~Lomholt,
``Open string fluctuations in AdS with and without torsion,''
Phys. Rev. D \textbf{68}, 066002 (2003)
doi:10.1103/PhysRevD.68.066002
[arXiv:hep-th/0305034 [hep-th]].
\bibitem{Larsen:1996yb}
A.~L.~Larsen and C.~O.~Lousto,
``On the stability of spherical membranes in curved space-times,''
Nucl. Phys. B \textbf{472}, 361-376 (1996)
doi:10.1016/0550-3213(96)00209-X
[arXiv:gr-qc/9602009 [gr-qc]].
\bibitem{Larsen:1995af}
A.~L.~Larsen and N.~G.~Sanchez,
``Strings and multistrings in black hole and cosmological space-times,''
NATO Sci. Ser. C \textbf{476}, 65-103 (1996)
[arXiv:hep-th/9504007 [hep-th]].

\bibitem{Mahapatra:1996ap}
S.~Mahapatra,
``String propagation in four-dimensional dyonic black hole background,''
Phys. Rev. D \textbf{55}, 6403-6408 (1997)
doi:10.1103/PhysRevD.55.6403
[arXiv:hep-th/9608028 [hep-th]].
\bibitem{Li:2018jxy}
A.~C.~Li, W.~L.~Xu and D.~F.~Zeng,
``Linear Stability Analysis of Evolving Thin Shell Wormholes,''
JCAP \textbf{03}, 016 (2019)
doi:10.1088/1475-7516/2019/03/016
[arXiv:1812.07224 [hep-th]].
\bibitem{Norma:2016soj}
B.~F.~Norma, C.~Cuauhtemoc, C.~Miguel and R.~Efrain,
``Covariant approach of perturbations in Lovelock type brane gravity,''
Class. Quant. Grav. \textbf{33}, no.24, 245012 (2016)
doi:10.1088/0264-9381/33/24/245012
[arXiv:1602.04892 [gr-qc]].
\bibitem{Hioki:2009hp}
K.~Hioki, U.~Miyamoto and M.~Nozawa,
``Stability of branes trapped by d-dimensional black holes,''
Phys. Rev. D \textbf{80}, 084011 (2009)
doi:10.1103/PhysRevD.80.084011
[arXiv:0908.1019 [hep-th]].


\bibitem{Brandenberger:2007by}
R.~Brandenberger, H.~Firouzjahi and O.~Saremi,
``Cosmological Perturbations on a Bouncing Brane,''
JCAP \textbf{11}, 028 (2007)
doi:10.1088/1475-7516/2007/11/028
[arXiv:0707.4181 [hep-th]].


\bibitem{Barik:2017opb}
S.~P.~Barik, K.~L.~Panigrahi and M.~Samal,
``Perturbation of pulsating strings,''
Eur. Phys. J. C \textbf{78} (2018) no.11, 882
doi:10.1140/epjc/s10052-018-6362-3
[arXiv:1708.05202 [hep-th]].

\bibitem{Capovilla:2021xfy}
R.~Capovilla, G.~Cruz and E.~Y.~L\'opez,
``Covariant higher order perturbations of branes in curved spacetime,''
Phys. Rev. D \textbf{105}, no.2, 025011 (2022)
doi:10.1103/PhysRevD.105.025011
[arXiv:2110.14894 [hep-th]].

\bibitem{Larsen:2000sg}
A.~L.~Larsen and A.~Nicolaidis,
``Second order perturbations of a macroscopic string: Covariant approach,''
Phys. Rev. D \textbf{63}, 125006 (2001)
doi:10.1103/PhysRevD.63.125006
[arXiv:hep-th/0009151 [hep-th]].


\bibitem{Kiosses:2014tua}
V.~Kiosses and A.~Nicolaidis,
``Second order perturbations of relativistic membranes in curved spacetime,''
Phys. Rev. D \textbf{89}, no.12, 124016 (2014)
doi:10.1103/PhysRevD.89.124016
[arXiv:1404.4166 [hep-th]].
\bibitem{deVega:1987veo}
H.~J.~de Vega and N.~G.~Sanchez,
``A New Approach to String Quantization in Curved Space-Times,''
Phys. Lett. B \textbf{197} (1987), 320-326
doi:10.1016/0370-2693(87)90392-3

\bibitem{deVega:1988ch}
H.~J.~de Vega and N.~G.~Sanchez,
``The Scattering of Strings by a Black Hole,''
Nucl. Phys. B \textbf{309} (1988), 577-590
doi:10.1016/0550-3213(88)90459-2

\bibitem{deVega:1988jh}
H.~J.~de Vega and N.~G.~Sanchez,
``Quantum Dynamics of Strings in Black Hole Space-times,''
Nucl. Phys. B \textbf{309} (1988), 552-576
doi:10.1016/0550-3213(88)90458-0

\bibitem{Delduc:2013qra}
Delduc, Francois and Magro, Marc and Vicedo, Benoit, "An integrable deformation of the $AdS_5 \times S^5$ superstring action",Phys. Rev. Lett
\textbf{112} (2014) no.5,051601
doi = 10.1103/PhysRevLett.112.051601
[arXiv: 1309.5850 [hep-th]].
\bibitem{Beisert:2008tw}
N.~Beisert and P.~Koroteev,
``Quantum Deformations of the One-Dimensional Hubbard Model,''
J. Phys. A \textbf{41} (2008), 255204
doi:10.1088/1751-8113/41/25/255204
[arXiv:0802.0777 [hep-th]].
\bibitem{Hoare:2023zti}
B.~Hoare, A.~L.~Retore and F.~K.~Seibold,
``Elliptic deformations of the $\mathsf{AdS}_3 \times \mathsf{S}^3 \times \mathsf{T}^4$ string,''
[arXiv:2312.14031 [hep-th]].
\bibitem{Hoare:2022asa}
B.~Hoare, F.~K.~Seibold and A.~A.~Tseytlin,
``Integrable supersymmetric deformations of AdS$_{3}$\texttimes{} S$^{3}$\texttimes{} T$^{4}$,''
JHEP \textbf{09}, 018 (2022)
doi:10.1007/JHEP09(2022)018
[arXiv:2206.12347 [hep-th]].
\bibitem{Hoare:2014oua}
B.~Hoare,
``Towards a two-parameter q-deformation of AdS$_3 \times S^3 \times M^4$ superstrings,''
Nucl. Phys. B \textbf{891}, 259-295 (2015)
doi:10.1016/j.nuclphysb.2014.12.012
[arXiv:1411.1266 [hep-th]].

\bibitem{Hoare:2014pna}
B.~Hoare, R.~Roiban and A.~A.~Tseytlin,
``On deformations of $AdS_n$ x $S^n$ supercosets,''
JHEP \textbf{06}, 002 (2014)
doi:10.1007/JHEP06(2014)002
[arXiv:1403.5517 [hep-th]].
\bibitem{Seibold:2021lju}
F.~K.~Seibold, S.~J.~van Tongeren and Y.~Zimmermann,
``On quantum deformations of AdS$_{3}$ \texttimes{} S$^{3}$ \texttimes{} T$^{4}$ and mirror duality,''
JHEP \textbf{09}, 110 (2021)
doi:10.1007/JHEP09(2021)110
[arXiv:2107.02564 [hep-th]].
\bibitem{Garcia:2021iox}
J.~M.~N.~Garc\'\i{}a and L.~Wyss,
``Three-parameter deformation of \ensuremath{\mathbb{R}} \texttimes{} S$^{3}$ in the Landau-Lifshitz limit,''
JHEP \textbf{07}, 028 (2021)
doi:10.1007/JHEP07(2021)028
[arXiv:2102.06419 [hep-th]].
\bibitem{Wen:2019mgv}
W.~Y.~Wen and S.~Kawamoto,
``Spin chains and classical strings in two parameters $q$-deformed AdS$_3\times$S$^3$,''
Chin. J. Phys. \textbf{64}, 348-362 (2020)
doi:10.1016/j.cjph.2019.12.020
[arXiv:1911.01567 [hep-th]].
\bibitem{Seibold:2019dvf}
F.~K.~Seibold,
``Two-parameter integrable deformations of the $AdS_3 \times S^3 \times T^4$ superstring,''
JHEP \textbf{10}, 049 (2019)
doi:10.1007/JHEP10(2019)049
[arXiv:1907.05430 [hep-th]].
\bibitem{Cherednik:1981df}
I.~V.~Cherednik,
``Relativistically Invariant Quasiclassical Limits of Integrable Two-dimensional Quantum Models,''
Theor. Math. Phys. \textbf{47}, 422-425 (1981)
doi:10.1007/BF01086395

\bibitem{Kameyama:2014vma}
T.~Kameyama and K.~Yoshida,
``A new coordinate system for $q$-deformed AdS$_{5} \times$ S$^5$ and classical string solutions,''
J. Phys. A \textbf{48}, no.7, 075401 (2015)
doi:10.1088/1751-8113/48/7/075401
[arXiv:1408.2189 [hep-th]].


\end{thebibliography}
\end{document}